\documentclass{ws-procs9x6}
\usepackage{epstopdf}

\def\etal {{\it et al.}}

\begin{document}

\title{PROBING PHYSICS BEYOND THE STANDARD MODEL WITH 
HE/XE CLOCK COMPARISON EXPERIMENTS}

\author{F.\ ALLMENDINGER and U.\ SCHMIDT$^*$}

\address{Physikalisches Institut, Ruprecht-Karls-Universit\"at Heidelberg\\
69120 Heidelberg, Germany\\
$^*$E-mail: ulrich.schmidt@physi.uni-heidelberg.de}

\author{W.\ HEIL, S.\ KARPUK, A.\ SCHARTH, 
Y.\ SOBOLEV, K.\ TULLNEY and S.\ ZIMMER}

\address{Institut f\"{u}r Physik, Johannes Gutenberg-Universit\"{a}t\\ 
55099 Mainz, Germany}

\begin{abstract}
The comparison of the free precession of co-located $^{3}$He-$^{129}$Xe 
spins (clock comparison) enables us to search for very tiny nonmagnetic 
spin interactions. With our setup we could establish new limits for 
Lorentz invariance violating interactions of spins with a 
relic background field which permeates the Universe and points in a 
preferred direction in space.
\end{abstract}

\bodymatter

\section{Experimental setup and principle of measurement}

In our experiments we employ polarized $^{3}$He and $^{129}$Xe as nuclear 
spin clocks. The two polarized gas species at pressure of order mbar are 
filled in a low-relaxation spherical glass cell with diameters between 6~cm 
and 10~cm together with nitrogen, which acts as buffer gas. The cell is 
positioned in the center of a homogeneous static magnetic field of about 
400~nT. The field is generated by means of Helmholtz coils, which are mounted 
inside the 7 layer magnetic shield of BMSR-2 at the PTB in Berlin. Details 
of the setup are given in Refs.\ \refcite{R1} and \refcite{R2}. At that field strength, the Larmor 
frequency of $^{3}$He is about $\omega _{He}\approx $2$\pi \cdot $13~Hz 
and $\omega _{Xe}\approx $2$\pi \cdot $4.7~Hz respectively. To measure 
these low precession frequencies, we use low-noise low-temperature DC-SQUID 
gradiometers as magnetic flux detectors. Due to the very low field gradients 
of order pT/cm at the location of our cell, we reached $T_2^\ast $ times of 
8h for $^{129}$Xe and up to 100h for $^{3}$He \cite{R3}. Therefore we can measure 
the free induction decay over a period $T$ of one day, thanks to a measured 
signal-to-noise ratio of SNR $>$ 3000:1 per $\sqrt{\text{Hz}}$. This long measurement period is 
important because the final statistical frequency error of our clock 
comparison experiments is proportional to $T^{-3/2}$ according to the 
Cramer Rao lower bound (CRLB)\cite{R2}.

To be sensitive to tiny nonmagnetic interactions, we must calculate the 
weighted difference of the respective Larmor frequencies of the 
co-located spin samples or the corresponding time integral, the Larmor 
phases, 
\begin{equation}
\label{eq1}
\Delta \omega =\omega_{\rm He} -\frac{\gamma_{\rm He} }{\gamma_{\rm Xe} }\omega _{\rm Xe} 
=0\quad ,\quad \Delta \Phi =\Phi_{\rm He} -\frac{\gamma_{\rm He} }{\gamma _{\rm Xe} 
}\Phi _{\rm Xe} =\mbox{const}.
\end{equation}
The weighting coefficient is $\gamma _{\rm He} /\gamma _{\rm Xe} 
=2.75408159(20)$. In doing so, magnetic field fluctuations are cancelled.
For the purpose of data analysis, we cut each run in subdata sets of 3.2~s 
time span. For each subdata set we extract both phase and amplitude of each 
spin species by fitting appropriate functions (for details, see Ref.\ \refcite{R4}). From these 
phase values we can extract the accumulated phases and with it the 
corresponding phase difference $\Delta \Phi (t)$. We reached with our last measurements in 2012 a 
phase sensitivity of $\Delta \Phi $=0.3~mrad or $\Delta $\textit{$\omega $~}=~2$\pi \cdot 
$40p~Hz. On a closer look, at that accuracy level, $\Delta \Phi $ is not a constant in time, as 
Eq.\ \eqref{eq1} may suggest. Instead higher order effects have to be take into account 
which can be parameterized as
\begin{eqnarray}
 \Delta \Phi (t)&=&
{\rm const} +a_{\rm lin} t+a_{\rm He} e^{-t/T_{\rm 2,He}^\ast 
}+a_{\rm Xe} e^{-t/T_{\rm 2,Xe}^\ast } 
\nonumber\\ 
&&
+b_{\rm He} e^{-2t/T_{\rm 2,He}^\ast }+b_{\rm Xe} e^{-2t/T_{\rm 2,Xe}^\ast }.
\label{eq2}
\end{eqnarray}
In Eq.\ \eqref{eq2}, the linear term $a_{\rm lin}$ has two contributions. 
One is trivial and stems from Earth's rotation. The other is caused by chemical 
shift, a correction for the finite density of the gases. The terms with 
$a_{He}$ and $a_{Xe}$ account for the Ramsey-Bloch-Siegert shift \cite{R5}. Each 
spin feels the magnetic moments of the other precessing spins of the same 
gas species that are slightly detuned in frequency due to magnetic field 
gradients (self-shift). Its value is proportional to the particular net 
magnetization of the spin species and decays therefore with the 
corresponding effective $T_{2}^{\ast }$ time of the free induction decay. 
Finally the last two terms of Eq.\ \eqref{eq2} take into account the shift due 
to the interaction of the spin species among each other (cross-talk). This 
interaction can be seen as a generalization of the well known Bloch-Siegert 
shift \cite{R6} and therefore its value is proportional to the magnetization 
squared. We determine the $T_{2}^{\ast }$ times independently by fitting 
the amplitude data of a run to an exponential. With the knowledge of both, 
the exact SQUID geometry and the geometry of the cell, we are also able to 
calculate from these fit results the $b$ coefficients of the Bloch-Siegert 
shift (see Eq.\ \eqref{eq2}). In contrast, we have no precise enough model for calculating 
the corresponding $a$ coefficients. As a consequence, we have to include the 
amplitudes of the linear term and the self-shift terms as free parameters 
while fitting the phase data. Now we consider a search for a finite 
nonmagnetic interaction. We include this interaction by adding an 
appropriate term $\Delta \Phi _{\rm nonmag}(t)$ to Eq.\ \eqref{eq2}. 
Depending on the time dependence of $\Delta \Phi _{\rm nonmag}(t)$, the 
amplitude of this term may be strongly correlated with the values of the $a$ 
coefficients of Eq.\ \eqref{eq2}, which we have also to include as free 
parameters into the fit, too. Therefore, careful correlation analysis of the 
fit parameters is mandatory. For high correlations, the sensitivity 
to nonmagnetic very weak interaction is reduced compared to the CRLB 
calculation \cite{R2}. 

\section{Constraints on a Lorentz invariance violating interaction}

In the context of the Standard-Model Extension (SME), possible interactions 
between the spin of a bound neutron and a relic background field are 
discussed. To determine the leading-order effects of a Lorentz violating 
potential $V$, it suffices to use a nonrelativistic description for the 
particles involved given by $V=\tilde {b}_{x,y,z}^n \cdot \vec {\sigma 
}_{x,y,z}^n $ \cite{R8}. We search for sidereal variations of the frequency of 
co-located spin species while the Earth and hence the laboratory 
reference frame rotates with respect to a preferred inertial frame. The 
observable to trace possible tiny sidereal frequency modulations is again 
the combination of measured Larmor frequencies (see Eq.\ (1)) and the weighted 
phase differences, respectively. Details of the setup and data analysis of 
our first run are given in Ref.\ \refcite{R4}. For the purpose of data analysis we had to 
add a sidereal modulation to the fitting function of Eq.\ \eqref{eq2} with 
\begin{equation}
\Delta \Phi _{\rm nonmag} (t)=
a_s \sin \left[ {\Omega _{SD} \left( 
{t-t_0 } \right)+\varphi _{SD} } \right]
-a_c \cos \left[ {\Omega _{SD} \left( {t-t_0 +\varphi _{SD} } 
\right)} \right].
\label{eq3}
\end{equation}
In this equation,
$\Omega _{SD }$ is the angular frequency of the sidereal day and $\varphi 
_{SD}$ represents the phase offset at the beginning $t_0$ of the first 
run. From that, the fit results for the amplitudes $a_{s}$ and $a_{c}$ as well 
as the RMS magnitude of the sidereal phase amplitude $\Phi _{SD} =\sqrt 
{a_s^2 +a_c^2 } $ could be extracted. In terms of the SME \cite{R8} we can express 
the sidereal phase amplitudes according to 
$a_{s(c)} =2\pi /\Omega _{SD} \cdot \upsilon _{X(Y)} $ 
with $2\pi \vert {\delta \upsilon _{X,Y} } 
\vert \hbar =\vert {2 \left( {1-\gamma _{He} /\gamma _{Xe} } 
\right) \sin \chi \, \tilde {b}_{X,Y}^n } \vert$, where $\chi $ is 
the angle between the Earth's rotation axis and the quantization axis of the 
spins. In Table \ref{tab} upper limits of the equatorial component 
$\tilde {b}_\bot ^n =\sqrt {(\tilde {b}_x^n )^2+(\tilde {b}_y^n )^2} $ of 
the background tensor field interacting with the spin of the bound neutron 
are given together with recent results from Ref.\ \refcite{R9}.

\begin{table}
\begin{center}
\tbl{Results for coefficients for Lorentz violation.}
{
\label{tab}
\begin{tabular}{|l|c|c|c|}
\hline
&&& Preliminary data,\\
Coefficient & 
Data 2009\cite{R4} & 
Romalis \etal\cite{R9} & 
March 2012 run \\
\hline
&&& \\
[-7pt]
$\tilde{b}_x^n \, \mathrm{[GeV]} \, (1\sigma)$& 
(3.4$\pm $1.7)$\cdot $10$^{-32}$& 
(0.1$\pm $1.6)$\cdot $10$^{-33}$& 
(4.1$\pm $4.7)$\cdot $10$^{-34}$ \\
[2pt]
$\tilde{b}_y^n $ [GeV] (1$\sigma )$& 
(1.4$\pm $1.3)$\cdot $10$^{-32}$& 
(2.5$\pm $1.6)$\cdot $10$^{-33}$& 
(2.9$\pm $6.2)$\cdot $10$^{-34}$ \\
[2pt]
$\tilde{b}_\bot ^n $ [GeV] (68{\%} C.L.)& 
$<\, 4.7\cdot 10^{-32}$& 
$<\, 3.7\cdot 10^{-33}$& 
$<\, 6.7\cdot 10^{-34}$ \\
[2pt]
$\tilde{b}_\bot ^n $ [GeV] (95{\%} C.L.)& 
$<\, 6.6\cdot 10^{-32}$& 
$<\, 5.5\cdot 10^{-33}$& 
$<\, 1.3\cdot 10^{-33}$ \\
[2pt]
\hline
\end{tabular}}
\end{center}
\end{table}

\vspace{-12pt}
\section{Conclusion}

Nuclear spin clocks, based on the detection of free spin precession of 
gaseous, nuclear polarized $^{3}$He or $^{129}$Xe samples with a SQUID as 
magnetic flux detector can be used as ultra-sensitive probe for nonmagnetic 
spin interactions, since the magnetic dipole interaction (Zeeman term) drops
out in case of co-located spin samples. With 
the long spin-coherence times, measurements of uninterrupted precession of 
$T \sim $ 1 day can be achieved at the present stage of investigation.
With an appropriate setup we also establish new limits for 
the pseudoscalar spin unpolarized matter interaction mediated by axion-like 
particles \cite{R7}. As the 
next challenging step, we want to employ this method for the search for an 
electric dipole moment of $^{129}$Xe.

\section*{Acknowledgments}

We thank our colleagues M. Burghoff, W. Kilian, S. Knappe-Gr\"uneberg, A. Schnabel, 
F. Seifert and L. Trahms from PTB in Berlin for their support during measurement time at PTB and for helpful discussions.

\end{document}